\documentclass[aps,preprintnumbers,superscriptaddress,showpacs,twocolumn]{revtex4}
\usepackage{epsfig}
\usepackage{psfrag}
\usepackage{amsfonts}
\usepackage{graphicx}
\usepackage{dcolumn}
\usepackage{bm}

\begin{document}

\title{Polarization energy loss  in hot viscous quark-gluon plasma}

\author{Bing-feng Jiang}
\email{jiangbf@iopp.ccnu.edu.cn} \affiliation{Center of Theoretical Physics and Department of Physics, Hubei Institute for Nationalities, Enshi, Hubei 445000, China }

\author{De-fu Hou}
\email{hdf@iopp.ccnu.edu.cn} \affiliation{Key Laboratory of Quark and Lepton Physics
(MOE) and Institute of Particle Physics, Central China Normal University, Wuhan 430079, China}

\author{Jia-rong Li}
\email{ljr@iopp.ccnu.edu.cn} \affiliation{Key Laboratory of Quark and Lepton Physics
(MOE) and Institute of Particle Physics, Central China Normal University, Wuhan 430079, China}

\date{\today}
\begin{abstract}
The gluon polarization tensor  for the quark-gluon plasma with shear viscosity is derived  with the viscous chromohydrodynamics. The longitudinal and transverse dielectric functions  are evaluated  from the gluon polarization tensor, through which the polarization energy loss suffered by a fast quark traveling through the viscous quark-gluon plasma is investigated.
The numerical analysis indicates that shear viscosity significantly reduces the polarization energy loss.

\end{abstract}

\maketitle
\section{Introduction}

There are two striking findings at the Relativistic Heavy-Ion Collider (RHIC). The first one is the strong jet quenching, which is believed to be a potential signal for the formation of the quark-gluon plasma(QGP). The other one is that the produced hot QCD plasma in heavy ion collisions behaves as a nearly perfect fluid with a small viscosity\cite{result1,result2,result3,result4}. The first results from the Large Hadron Collider (LHC) also qualitatively support the similar conclusions as seen at the RHIC\cite{alice,atlas}.

High transverse momentum partons produced by the hard scatterings in the initial stage in heavy ion collisions will loss energy by collisions with partons in medium or by bremsstrahlung of gluons and degrade, which is the so-called jet quenching.  Bjorken has firstly anticipated the jet quenching in QCD plasma by calculating the energy loss suffered by the fast parton by considering binary elastic scatterings of it off thermal quarks and gluons in thermal bath\cite{bjorken}. The resulting collisional energy loss for massless quark is
\begin{equation}\label{bjorken}
-\frac{dE}{dx}=\frac{8}{3}\pi \alpha_s^2 T^2(1+\frac{n_f}{6})\log\frac{q_{max}}{q_{min}}.
\end{equation}
That result is infrared and ultraviolet divergences which are regulated by introducing a reasonable minimum momentum transfer $q_{min}$ about Debye mass scale by hand and a maximum momentum transfer $q_{max}$ according to the scattering kinematics, respectively.
The amount of quenching is  relevant to the state of matter of fireball produced in heavy ion collisions, ie, the QGP or the hot hadron matter. Therefore, the  jet quenching might be severed as a probe for the QGP formation in heavy ion collisions\cite{bjorken,gyulassy}.

In light of the technique of plasma physics\cite{ichimaru}, Thoma and  Gyulassy have developed another formula for collisional energy loss  by considering the reaction of the color Lorentz
force of the chromoelectric field induced by the fast quark  on itself\cite{thoma91a}.
The formula relates the energy loss to the longitudinal and transverse dielectric functions
which can be derived in terms of the gluon self-energy, as shown in Eq.(\ref{enlossm}) in the next section. One advantage of that approach is that the infrared divergence is automatically cut off by the plasma collective effect.  However, it is argued that that approach  is not applicable for the large momentum transfer, because  it is meaningless to consider an induced chromoelectric field with a wavelength shorter than the average distance between plasma partons\cite{mrowczynski}. In order to circumvent the problem, some people have divided the collisional energy loss into two parts.  The first part is for the soft momentum transfer relating to the plasma polarization effect, ie, the polarization energy loss, and the other is for the hard momentum transfer due to individual collisions between the fast quark and plasma partons. The approach of Thoma and Gyulassy is applicable for the polarization energy loss,  while the Bjorken's formula can be applied to the latter. The full energy loss is the sum of the two  parts\cite{mrowczynski,thoma91b}.
But there is still a problem that the full expression of the energy loss is
dependent on the intermediate momentum scale relating to the separation of soft and hard momenta. Latter,   Braaten and Thoma have constructed a systematic field theory method to deal with the
collisional energy loss including both plasma polarization effect and individual collisions
between partons. The hard thermal loop resummation approach (HTLR)\cite{htlr} should be used
for the soft momentum transfer, while naive perturbative theory is applied for the hard momentum
transfer. The dependence of result on the intermediate momentum scale is
removed\cite{braaten91a,braaten91b}.

The subsequent studies show that the induced  radiative energy loss dominates the collisional one and might be the main mechanism for the jet quenching, for reviews to see Refs.\cite{baier,loss1,loss2,loss3,loss5,loss6}. The suppression of the light hadron spectra at large $p_T$ observed at the RHIC can be interpreted qualitatively due to radiative energy loss suffered by its parent partons before hadronization.
However, the spectrum of non-photonic electrons\cite{electron} which are primarily produced due
to semi-leptonic decay of heavy quarks shows that it is not sufficient to explain the heavy
quark quenching by considering radiative energy loss merely.
That fact has attracted  people to rescrutinize the energy loss mechanism\cite{qin08,cao,wicks07a,zakharov,mazumder05,mustafa05b,mustafa05a,meistrenko,peshier12,gossiaux08,zapp,gossiaux09,moore05}. Some recent studies show that  the collisional energy loss is comparable to the radiative one or even larger  than the latter in some energy
region\cite{mazumder05,mustafa05b,mustafa05a,meistrenko,peshier12,gossiaux08,zapp,gossiaux09,moore05}.
In addition, the study of the collisional energy loss has been extended to hot anisotropic QED and QCD mediums\cite{romatschke04,romatschke05}. An operator definition and derivation of collisional energy and momentum loss in relativistic plasmas was recently proposed \cite{xing14}.

According to the Refs.\cite{groot,arnold,teaney,dusling12},
viscosity will modify the distribution functions of the constituents  of the QGP. Therefore it will affect the the gluon self-energy through which the dielectric functions will  be modified.  In terms of  Eq.(\ref{enlossm}) in the next section, viscosity will have an impact on the polarization energy loss. In the present paper, by following the theoretic framework of Thoma and Gyulassy \cite{thoma91a}, we will study the effect of shear viscosity on the polarization energy loss suffered by a fast quark traveling through the QGP.

It is argued  that chromohydrodynamics can describe
the polarization effect as the kinetic theory~\cite{manuel1}. In  recent
papers\cite{jiang1,jiang2},  the authors have extended the ideal chromohydrodynamics~\cite{manuel2,manuel3} to the viscous one in terms of the QGP kinetic theory and the distribution function modified by the shear viscosity. Under that framework, the gluon polarization tensor is derived, through which the longitudinal dielectric function and the refraction index in the viscous QGP have been studied in \cite{jiang1} and \cite{jiang2}, respectively. In addition, based on the longitudinal dielectric function, the induced color charge distribution\cite{jiang3} and the corresponding wake potential\cite{jiang4} induced by the fast parton traveling through the viscous QGP have been investigated later.

In the present paper, by following the  gluon polarization tensor derived from the viscous chromohydrodynamics, we will evaluate  the  longitudinal and transverse dielectric functions in the QGP associated with shear viscosity. Then, based on them, we will study the polarization energy loss suffered by the fast parton traveling through the viscous QGP.

The paper is organized as follows. In Section 2, we will briefly review the framework of Thoma and Gyulassy  for the polarization energy loss  suffered by the fast parton traversing the  QGP. Then, we will derive the longitudinal and transverse dielectric functions from the polarization tensor obtained with the viscous chromohydrodynamics. In Section 3, based on the derived dielectric functions, we will evaluate the polarization energy loss and discuss the viscous effect on it.  Section 4 is summary. We will give an alternative derivation of the transverse dielectric function in Appendix A and make a detailed derivation of $\rm Im \varepsilon_L(\omega,k)|_{\omega=\textbf{k}\cdot\textbf{v}}$ in Appendix B.

The natural units $k_B=\hbar=c=1$, the metric $g_{\mu\nu}=(+,-,-,-)$ and  the notations $K=(\omega,\textbf{k})$, $k=|\textbf{k}|$ are used in the paper.

\section{The polarization energy loss in the viscous quark-gluon plasma}
In this section, at first, we will  briefly review the framework  of polarization energy loss in Ref.\cite{thoma91a}. The longitudinal and transverse dielectric functions play a critical role in  the polarization energy loss. Then, according to the gluon polarization tensor obtained from viscous chromohydrodynamics, we will derive the longitudinal and transverse dielectric functions in the presence of shear viscosity, through which we can study the viscous effect on the polarization energy loss suffered by the fast parton traversing the viscous QGP.

\subsection{The formula for polarization energy loss}
When the fast quark is introduced into the QGP, a chromoelectric field is induced in medium. The color Lorentz force due to the induced chromoelectric field will exert in return on the fast quark itself, which will cause the energy loss to the fast quark. The formula of energy loss is given by\cite{thoma91a}
\begin{equation}\label{enloss}
-\frac{dE}{dx}=\frac{\textbf{v}}{v} q^a \cdot \rm Re \textbf{E}_{ind}^a(\textbf{x}=\textbf{v}t,t),
\end{equation}
where $\textbf{E}_{ind}^a(\textbf{x}=\textbf{v}t,t)$ is the induced chromoelectric field and $v=|\textbf{v}|$.   $q^a$ is the color charge relating to the fast quark and defined as  $q^aq^a=C_F\alpha_s$ with strong coupling constant $\alpha_s=g^2/4\pi$ and Casimir invariant $C_F=4/3$ for the fundamental representation. We assume that the QGP is a static medium and the running coupling constant is invariant.

Due to the external current of the fast quark $\textbf{j}^a_{ext}$,  a chromoelectric field will be induced in the medium. In the linear response theory, the total chromoelectric field can be expressed in the momentum space as\cite{thoma91a}
\begin{equation}\label{tote}
[\varepsilon_{ij}(\omega,k)-\frac{k^2}{\omega^2}(\delta_{ij}-\frac{k_ik_j}{k^2})]
\textbf{E}^a_{tot}(\omega,k)=\frac{4\pi}{i \omega}\textbf{j}^a_{ext}(\omega,k).
\end{equation}
Here, $\varepsilon_{ij}(\omega,k)$ is the dielectric tensor which reflects the chromoelectromagnetic properties of the QGP medium. In the isotropic and homogeneous medium, $\varepsilon_{ij}(\omega,k)$ can be decomposed into two components
\begin{equation}\label{diet}
\varepsilon_{ij}(\omega,k)=\varepsilon_L(\omega,k)\frac{k_ik_j}{k^2}+
\varepsilon_T (\delta_{ij}-\frac{k_ik_j}{k^2}),
\end{equation}
ie, longitudinal and transverse dielectric functions $\varepsilon_L(\omega,k)$, $\varepsilon_T(\omega,k)$.

The external current due to the fast quark  in the momentum space can be denoted as\cite{thoma91a}
\begin{equation}\label{curr}
\textbf{j}^a_{ext}(\omega,k)=2\pi q^a \textbf{v} \delta(\omega-\textbf{k}\cdot \textbf{v}).
\end{equation}
According to Eqs.(\ref{tote})(\ref{diet})(\ref{curr}), by working out the induced chromoelectric field and substituting  it into Eq.(\ref{enloss}), we will arrive the  polarization energy loss\cite{thoma91a,koike}
\begin{eqnarray}\label{enlossm}
-\frac{dE}{dx}=&-&\frac{C_F\alpha_s}{2\pi^2v} \int d^3k \{\frac{\omega}{k^2}[ \rm Im \varepsilon_L^{-1}\nonumber\\ &+&
(v^2k^2-\omega^2) \rm Im (\omega^2\varepsilon_T-k^2)^{-1}]\}_{\omega=\textbf{k}\cdot \textbf{v}}.
\end{eqnarray}
It can be seen that the longitudinal and transverse dielectric functions dominate the polarization energy loss. If the viscous information of the QGP can be embedded in the longitudinal and transverse dielectric functions, one can study the viscous effect on the polarization energy loss.

\subsection{The dielectric functions in viscous quark-gluon plasma}
Viscosity will modify the distribution functions of the constituents of a microscopic system\cite{groot,arnold,teaney,dusling12}.
If only shear viscosity is taken into account,
the modified distribution function can be written as\cite{groot,arnold,teaney,dusling12,hou,jeon,duslingprc}
\begin{equation}
Q=Q_o+\delta Q=Q_o+\frac{c'}{2T^3}\frac{\eta}{s}Q_o(1\pm Q_o)p^\mu
p^\nu\langle \nabla_\mu u_\nu \rangle. \label{dis}
\end{equation}
In Eq.(\ref{dis}), ``$+$'' is for boson, while ``$-$'' is for
fermion.
 $c'=\pi^4/90\zeta(5)$ and $c'=14
\pi^4/1350\zeta(5)$ are for massless boson~\cite{duslingprc, teaney}
and massless fermion~\cite{duslingnpa} respectively. $\langle
\nabla_\mu u_\nu \rangle = \nabla_\mu u_\nu + \nabla_\nu u_\mu -
\frac{2}{3} \Delta_{\mu\nu}\nabla_{\gamma}u^{\gamma}$, $\nabla_{\mu}
= (g_{\mu\nu} - u_{\mu}u_{\nu})\partial^{\nu}$, $\Delta^{\mu\nu}=
g^{\mu\nu} - u^{\mu}u^{\nu}$;   $\eta, s$,   $T$, $Q_o$
represent the  shear viscosity, the entropy density,  the temperature of the system and the ideal distribution function of boson or fermion.

It is very difficult  to evaluate the gluon self-energy with the QGP kinetic theory associated with the  distribution function modified by shear viscosity Eq.(\ref{dis}) (We have noticed that in a very recent literature the  vector meson thermal self-energy in a hot viscous hadronic medium has been addressed in terms of the  distribution function modified by shear viscosity and the forward scattering amplitude\cite{vujanovic}). Fortunately, the fluid equations are rather simpler than the kinetic theory and usually used to study the plasma properties. In light of Refs.~\cite{manuel2,manuel3}, we have derived the viscous chromohydrodynamic equations  by expanding the non-Abelian kinetic equations in momentum moments  and truncating the expansion at the second moment level with the modified distribution function (\ref{dis}) ~\cite{jiang1,jiang2}.

The viscous chromohydrodynamic equations can be linearized around
the stationary, colorless and homogeneous plasma state which is
described by $\bar{n}$,$\bar{u}^\mu$,$\bar{p}$ and $\bar{\epsilon}$.
By using an  equation of state~(EoS)  $\delta
p_a=c_s^2\delta\epsilon_a$, one can obtain the colored fluctuations
of these hydrodynamic quantities $\delta n_a$,$\delta
u^\mu_a$,$\delta p_a$ and $\delta \epsilon_a $. In terms of the derived
fluctuations, one can obtain the fluctuation of the color current  $\delta j^\mu_a(\omega,k)$. Then, according to the
relation between the color current and the gauge field in the linear
response theory $\delta
j^\mu_a(\omega,k)=-\Pi^{\mu\nu}_{ab}(\omega,k)A_{\nu,b}(\omega,k)$,  one can extract the polarization tensor $\Pi^{\mu\nu}_{ab}(\omega,k)$~\cite{jiang1,jiang2}
\begin{eqnarray}
\Pi_{ab}^{\mu\nu}(\omega,k)&=&-\delta_{ab} \{\omega_p^2
 \cdot\frac{1}{1+D(K^2-(K\cdot \bar{u})^2)}\cdot\frac{1}{(K\cdot
\bar{u})^2}\nonumber\\&\cdot&[(K\cdot \bar{u})(\bar{u}^\mu k^\nu+k^\mu
\bar{u}^\nu)-K^2\bar{u}^\mu \bar{u}^\nu\nonumber\\&-&(K\cdot
\bar{u})^2g^{\mu\nu}+(B+E)\cdot[K^2(K\cdot \bar{u})(\bar{u}^\mu
k^\nu\nonumber\\&+&k^\mu\bar{u}^\nu)-k^\mu k^\nu (K\cdot
\bar{u})^2-K^4\bar{u}^\mu \bar{u}^\nu]]\},\label{pol}
\end{eqnarray}
where $\omega^2_p=\frac{g^2\bar{n}^2}{2(\bar{\epsilon}+\bar{p})}$ is the square of the plasma frequency and
\begin{eqnarray}\label{para}
 B&=&-\frac{c_s^2}
  {\omega^2-c_s^2k^2},\ \ \ \ \ \  \ \ \ \ \ \  D=\frac{\eta}{sT\omega},\nonumber\\
  E
 &=&-\frac{\frac{\eta\omega}{sT}
 (1+4\frac{c_s^2k^2}
  {\omega^2-c_s^2k^2})}
  {3\omega^2-3c_s^2k^2-
 4\frac{\eta\omega k^2}
{sT}}.
\end{eqnarray}
We have briefly reviewed the polarization tensor derived from the viscous chromohydrodynamics, for details please refer to Refs.\cite{jiang1,jiang2}.

The dielectric tensor relates to the gluon polarization tensor as
\begin{equation}\label{diet2}
\varepsilon^{ij}(\omega,k)=\delta^{ij}+\frac{\Pi^{ij}(\omega,k)}{\omega^2}.
\end{equation}
According to Eq.(\ref{pol}), we can obtain
\begin{eqnarray}\label{pij}
\Pi^{ij}(\omega,k)=\frac{\omega_p^2}{1-Dk^2}\{g^{ij}+(B+E)k^i k^j\}.
\end{eqnarray}
Therefore, the dielectric tensor can be expressed as
\begin{equation}\label{diet3}
\varepsilon^{ij}(\omega,k)=\delta^{ij}+\frac{\omega_p^2}{\omega^2}\frac{1}{1-Dk^2}\{g^{ij}+(B+E)k^i k^j\}.
\end{equation}

In terms of Eqs.(\ref{diet3}) and (\ref{diet}), one can obtain the longitudinal and transverse dielectric functions
\begin{eqnarray}\label{dl}
\varepsilon_L(\omega,k)&=&\frac{k_ik_j}{k^2}\varepsilon^{ij}(\omega,k)\nonumber\\
&=&1+\frac{\omega_p^2}
{\omega^2}\frac{1}{1-Dk^2}\{\frac{k_ik_j}{k^2}g^{ij}+\frac{k_ik_j}{k^2}(B+E)k^i k^j\}
\nonumber\\&=&1-\frac{\omega_p^2}
{\omega^2}\frac{1}{1-Dk^2}\{1-(B+E)k^2\},
\end{eqnarray}
\begin{eqnarray}\label{dt}
\varepsilon_T(\omega,k)&=&\frac{1}{2}(\delta_{ij}-\frac{k_ik_j}{k^2})\varepsilon^{ij}(\omega,k)
=\frac{1}{2}(\delta_{ij}-\frac{k_ik_j}{k^2})\nonumber\\&\times&(\delta^{ij}+\frac{\omega_p^2}
{\omega^2}\frac{1}{1-Dk^2}\{g^{ij}+(B+E)k^i k^j\})\nonumber\\&=&
1-\frac{\omega_p^2}{\omega^2}\frac{1}{1-Dk^2}.
\end{eqnarray}
Substituting $B$,$D$, $E$ in Eq.(\ref{para}) and the effective sound speed $c_s=\sqrt{\frac{1}{3(1+\frac{1}{2y}\log\frac{1-y}{1+y})}+\frac{1}{y^2}}$
$(y=\frac{k}{\omega})$ \cite{manuel1,jiang1,jiang2} into (\ref{dl})(\ref{dt}), we can arrive the longitudinal  dielectric function\cite{jiang1,jiang2}
\begin{eqnarray}\label{dfl}
 & &\varepsilon_L(\omega,k)
=1+\frac{3\omega_p^2}{k^2}[1-\frac{\omega}{2k}
 (\ln|\frac{\omega+k}{\omega-k}|-i\pi \nonumber\\&\cdot&\Theta(k^2-\omega^2))]
-\frac{12\omega_p^2}
 {k^2}\frac{\eta\omega}{sT}\times
 \{1-\frac{\omega}{k}
 \ln|\frac{\omega+k}{\omega-k}|
 \nonumber\\&+&\frac{\omega^2}{4k^2}
 (\ln|\frac{\omega+k}{\omega-k}|)^2-
 \frac{\omega^2}{4k^2}\pi^2\Theta(k^2-\omega^2)\nonumber\\&+& i
 (\frac{\omega}{k}\pi-\frac{\omega^2}{2k^2}\pi
 \ln|\frac{\omega+k}{\omega-k}|)\Theta(k^2-\omega^2)\}, \label{vdief}
\end{eqnarray}
and the transverse one
\begin{equation}\label{dft}
\varepsilon_T(\omega,k)=1-\frac{\omega_p^2}{\omega^2}\frac{1}
{1-\frac{\eta}{s}\frac{k^2}{\omega T}}.
\end{equation}


Through the polarization tensor with the viscous chromohydrodynamic approach, shear viscosity embeds in the longitudinal and transverse dielectric functions. Substituting the dielectric functions Eqs.~(\ref{dfl})(\ref{dft}) into Eq.~(\ref{enlossm}), one can study
the polarization energy loss of the fast quark in the QGP in the presence of  shear viscosity.

\section {Numerical analysis}

From the discussion in the section II, one can see that the dielectric functions dominate the polarization energy loss. Therefore, the detailed investigation on the viscous dielectric functions  will shed light on the viscous effect on the polarization energy loss.
\begin{figure}
\begin{minipage}[h]{0.48\textwidth}
\centering{\includegraphics[width=8cm,height=4.944cm] {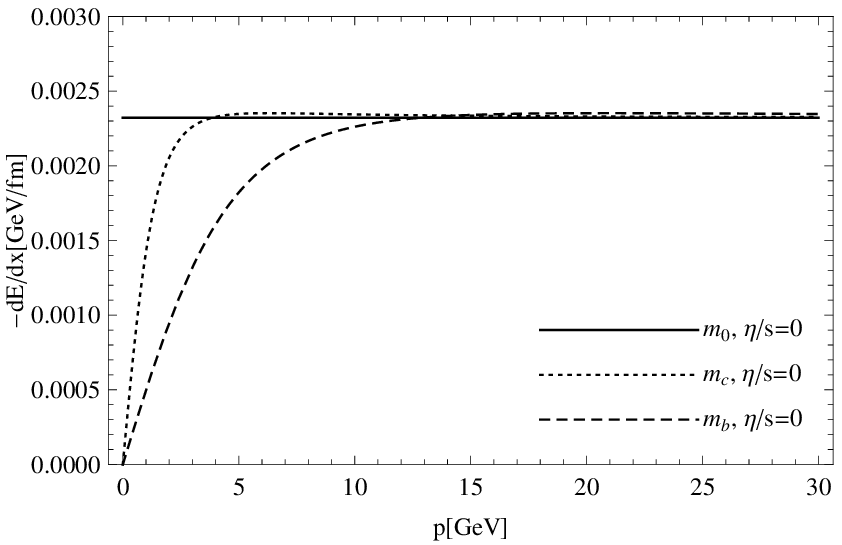}}
\end{minipage}
\begin{minipage}[h]{0.48\textwidth}
\centering{\includegraphics[width=8cm,height=4.944cm] {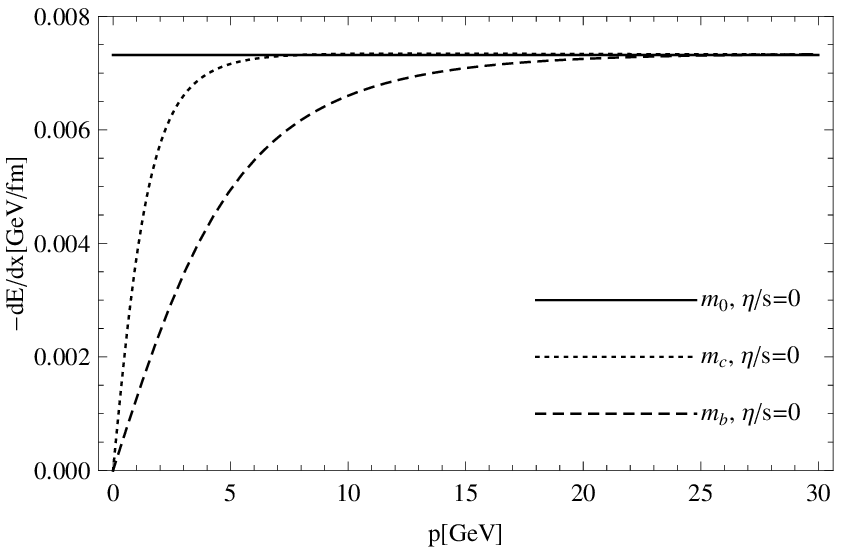}}
\end{minipage}
\caption{(color online)Polarization energy loss of a light, charm and bottom quark for $\eta/s=0.0$. The solid, dotted and dashed curves are for light, charm and bottom quark, respectively. Top panel: $k_{max}=0.5$GeV; Bottom panel: $k_{max}=0.7$GeV.}\label{v0}
\end{figure}

The same result of the longitudinal dielectric function (\ref{dfl}) has been obtained
with nonlinear viscous chromohydrodynamics which is derived from the non-Abelian kinetic theory and the dissipative distribution function determined by entropy production principle (EPP) \cite{ramos}. When $\eta/s=0$, $\varepsilon_L(\omega,k)$ recovers the result in the hard thermal loop approximation(HTLA)  obtained by the kinetic theory or finite temperature field theory\cite{weldon,heinz1,heinz,kapusta,bellac} if such relation $m^2_D=3\omega^2_p$ is adopted\cite{mustafa06}. In addition, the authors in the literature\cite{ramos} have reviewed that the longitudinal dielectric function (\ref{dfl}) might be applied  to study the energy loss.

On the other hand, the transverse dielectric function (\ref{dft}) is a pure real function of $\omega,k$ (which can be derived with an alternative way in Appendix A).  According to  formula (\ref{enlossm}), it indicates that the transverse dielectric function derived from the viscous chromohydrodynamics has no contribution to polarization energy loss. That result is reminiscent of the argument that in the HTLA the contribution to the polarization energy loss from the transverse dielectric function is much smaller than that from the longitudinal one and the polarization energy loss  mainly comes from the longitudinal dielectric function
\cite{mrowczynski,koike}.


\begin{figure}
\begin{minipage}[h]{0.48\textwidth}
\centering{\includegraphics[width=8cm,height=4.944cm] {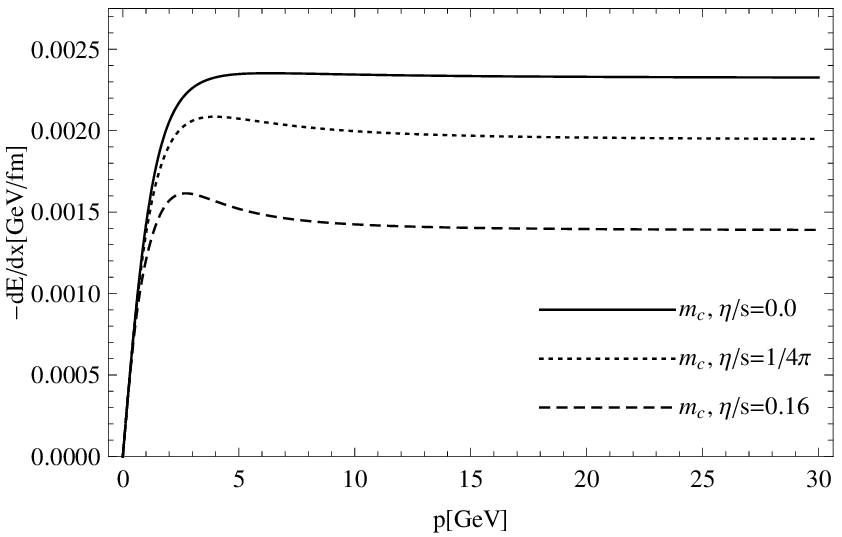}}
\end{minipage}
\begin{minipage}[h]{0.48\textwidth}
\centering{\includegraphics[width=8cm,height=4.944cm] {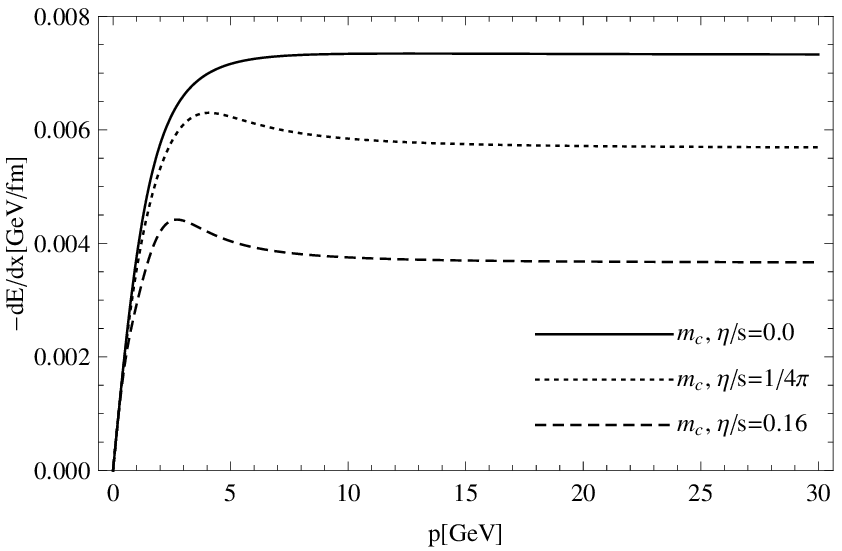}}
\end{minipage}
\caption{(color online)Polarization energy loss of a charm quark for different shear viscosity. The solid, dotted and dashed curves are for $\eta/s=0.0$, $\eta/s=1/4\pi$ and $\eta/s=0.16$, respectively. Top panel: $k_{max}=0.5$GeV; Bottom panel: $k_{max}=0.7$GeV.}\label{mc}
\end{figure}

\begin{figure}
\begin{minipage}[h]{0.48\textwidth}
\centering{\includegraphics[width=8cm,height=4.944cm] {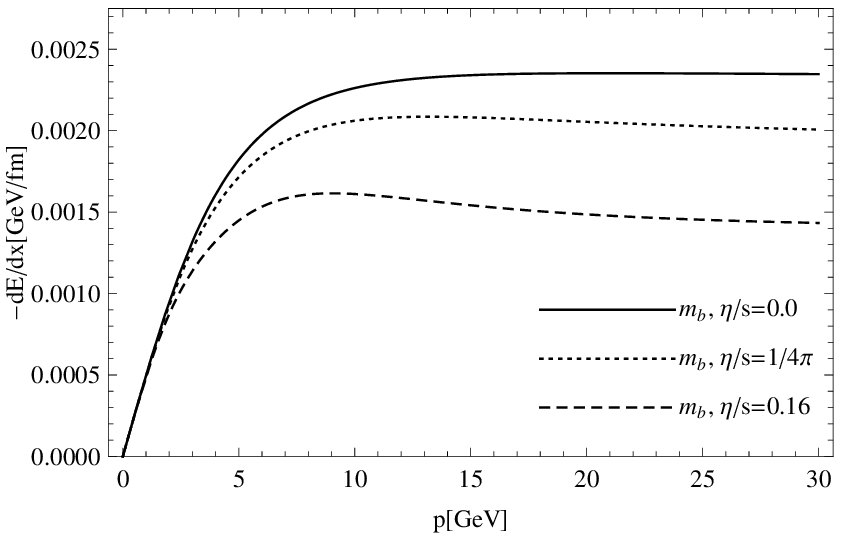}}
\end{minipage}
\begin{minipage}[h]{0.48\textwidth}
\centering{\includegraphics[width=8cm,height=4.944cm] {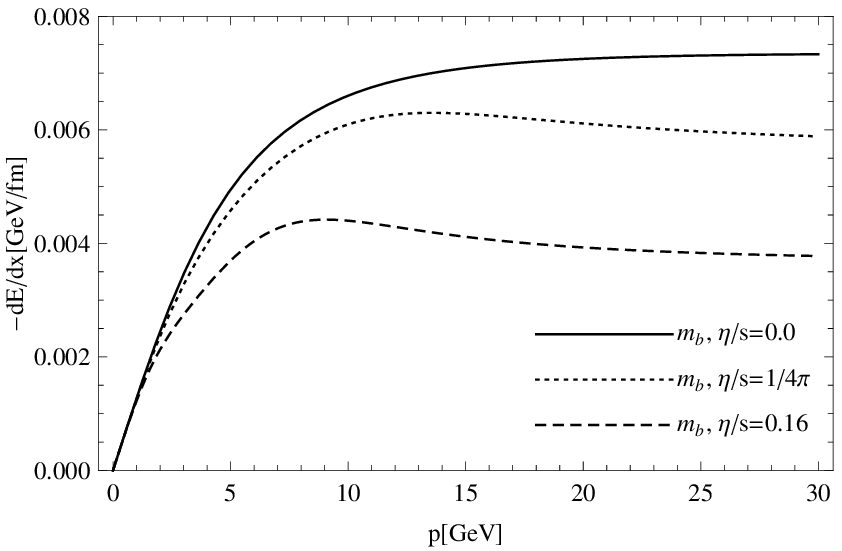}}
\end{minipage}
\caption{(color online)Polarization energy loss of a bottom quark for different shear viscosity. The solid, dotted and dashed curves are for $\eta/s=0.0$, $\eta/s=1/4\pi$ and $\eta/s=0.16$, respectively. Top panel: $k_{max}=0.5$GeV; Bottom panel: $k_{max}=0.7$GeV.}\label{mb}
\end{figure}

According to Eqs.~(\ref{dfl}) (\ref{dft}) and (\ref{enlossm}), we can get
\begin{eqnarray}\label{vloss}
-\frac{dE}{dx}&=&-\frac{C_F\alpha_s}{2\pi^2 v}\int d^3k
\frac{\omega}{k^2} \rm Im \varepsilon_L^{-1}(\omega,k)|_{\omega=\textbf{k}\cdot\textbf{v}}.
\end{eqnarray}
In terms of Eq.(\ref{dfl}), $\rm Im \varepsilon_L^{-1}(\omega,k)|_{\omega=\textbf{k}\cdot\textbf{v}}$ in the viscous QGP can be derived as (For detailed derivation, please prefer to Appendix B)
\begin{eqnarray}
& &\rm Im \varepsilon_L^{-1}(\omega,k)|_{\omega=\textbf{k}\cdot\textbf{v}}=-\frac{m_D^2\pi v \cos\theta}{2}\nonumber\\ &\cdot&\frac{k^2-w_1\cdot k^3}{[k^2+w_3\cdot k+w_2]^2+w_4\cdot[1-w_1\cdot k]^2},\label{18}
\end{eqnarray}
with
\begin{eqnarray}\label{parameter}
&w_1&=\frac{8v\cos\theta}
 {T}\frac{\eta}{s}(1-\frac{v\cos\theta}{2}
 \ln|\frac{1+v\cos\theta}{1-v\cos\theta}|),\nonumber\\
 &w_2&=m_D^2(1-\frac{v\cos\theta}{2}
 (\ln|\frac{1+v\cos\theta}{1-v\cos\theta}|)),\nonumber\\
 &w_3&=-\frac{4m_D^2v\cos\theta}
 {T}\frac{\eta}{s}\times
 \{1-v\cos\theta
 \ln|\frac{1+v\cos\theta}{1-v\cos\theta}|
\nonumber\\&+&\frac{v^2\cos^2\theta}{4}
 (\ln|\frac{1+v\cos\theta}{1-v\cos\theta}|)^2-
 \frac{v^2\cos^2\theta}{4}\pi^2\},
 \nonumber\\
 &w_4&=\frac{m_D^4\pi^2 v^2 \cos^2\theta}{4},
 \end{eqnarray}
where $\cos\theta=\textbf{k}\cdot\textbf{v}/{kv}$. For incident quark, $v=\frac{p}{\sqrt{p^2+M^2}}$\cite{thoma91a} where $M$ is  quark mass and $p$ is momentum.
Therefore, (\ref{vloss}) will turn to be
\begin{eqnarray}\label{vlossm}
&-&\frac{dE}{dx}=
\frac{m_D^2 C_F\alpha_sv}{2} \int_{-1}^1d(\cos\theta)\cos^2\theta \int_0^{k_{max}} dk \nonumber\\&\cdot&\{\frac{k^3-w_1\cdot k^4}{[k^2+w_3\cdot k+w_2]^2+w_4\cdot[1-w_1\cdot k]^2}\}.
\end{eqnarray}

In this paper, we  regard  shear viscous coefficient  as an input parameter to study its effect on the polarization energy loss.  The casual viscous hydrodynamics simulation fits well to experimental data on the  integrated elliptic flow coefficient $v_2$ at the RHIC with  $\eta/s\sim 0.16$\cite{romatschke2}, which is about two times of the famous bound result $\eta/s=\frac{1}{4\pi}$ of the strongly couple conformal field theory determined by the AdS/CFT correspondence\cite{bound}. Numerical results of the polarization energy loss are presented with those explicit values of
$\eta/s$.

In numerical analysis, we have assumed that quark mass $m_0=0$, $m_c=1.5$GeV  and $m_b=5.0$GeV for light, charm and bottom quark,
respectively. In addition, $T=0.3$GeV, $\alpha_s=0.3$ and $N_f=2$   are adopted. Under that condition, the Debye screening mass $m_D\sim0.7$GeV. The upper limit $k_{max}$ in the integration of the polarization energy loss (\ref{vlossm}) is a quantity of order of Debye screening mass. In numerical calculation, we choose two values  $k_{max}=0.5$GeV and $0.7$GeV for comparison.

When $\eta/s=0$, $w_1=w_3=0$, the longitudinal dielectric function (\ref{dfl}) recovers the result in the HTLA.  (\ref{vlossm}) turns to the polarization energy loss coming from the longitudinal dielectric function in the HTLA\cite{thoma91a,koike}.
We present the polarization energy loss in Fig.\ref{v0} for $\eta/s=0$ for light,
charm and bottom quarks. In Fig.\ref{v0}, the solid, dotted and dashed curves are for light, charm and bottom quarks, respectively.
The polarization energy loss curves for heavy quark resemble the one of collisional energy loss  obtained in Refs.\cite{thoma91a,braaten91a,braaten91b}. For light quark, $m_0=0$, thus $v=1$,
the polarization energy loss in (\ref{vlossm}) is independent on the momentum of the
incident quark, as shown of the solid curve in Fig.\ref{v0}.
That conclusion is implied in Ref.\cite{koike} where the polarization energy loss is expressed
as a function of the velocity of the incident quark. In addition, it is shown that polarization
energy loss for $k_{max}=0.7$GeV is much larger than that for $k_{max}=0.5$GeV. That result indicates that the polarization energy loss is sensitive to $k_{max}$ which
is consistent with the argument in \cite{koike}. Nevertheless,
it can be seen in Fig.\ref{v0}, the following Fig.\ref{mc} and Fig.\ref{mb} that the energy loss
shows the similar viscous and momentum-dependent behavior for $k_{max}=0.5$GeV and  $k_{max}=0.7$GeV.

In Fig.\ref{mc} and Fig.\ref{mb}, we demonstrate the polarization energy loss of  charm and bottom quarks with different shear viscosity, respectively. The solid, dotted and dashed curves are for $\eta/s=0$, $\eta/s=1/4\pi$ and $\eta/s=0.16$, respectively.
The energy loss is a complex function of shear viscosity, incident quark momentum and  mass. At small incident momentum, shear viscosity has trivial effect on the polarization energy loss.  As the increase of the incident momentum, shear viscosity significantly reduces the polarization energy loss.  When $p\sim5$GeV for charm quark and $p\sim10$GeV for bottom quark, viscous curves become asymptotically horizontal  lines.  On the other hand, for a definite incident quark momentum, as shear viscosity increases, the polarization energy loss becomes even smaller. It should be noted that although the energy loss has been not addressed in detail in a recent literature\cite{ramos}, the authors have anticipated that shear viscosity will lead to a reduction of the energy loss according to the viscous behavior of the  longitudinal dielectric function.



The significant reduction of the energy loss in the presence of shear viscosity can retrospect to the second term  $w_1\cdot k^4$ in numerator in Eq. (\ref{vlossm}) which relates to the viscous term in the dielectric function Eq. (\ref{dfl}). The denominator of Eq. (\ref{vlossm}) is of the order $\sim k^4$, the first term of the numerator is of the order $\sim k^3$, therefore, the integration of $k$ for the first term  in Eq. (\ref{vlossm}) gives the result of order $\sim \ln k$ which is qualitatively consistent with the HTL result\cite{bjorken,thoma91a,braaten91a,braaten91b}.  The numerator and denominator of  the second term in Eq. (\ref{vlossm})  are the same  order of $\sim k^4$. As a result, the integration of $k$ for the second term gives a contribution of the order $\sim k$, which may lead to a substantial reduction of the polarization energy loss.

If we resort to the  energy loss in a more general collisional kinematic formula,
$-\frac{dE}{dx}$ is   proportionate  to the cross section $\sigma$ \cite{peshier06prl}
which is related to the mean free path $\lambda$ as $\sigma\propto\frac{1}{\lambda}$.
While shear viscosity is given by an expression of the form in terms of   $\lambda$
as $\eta\propto\lambda$\cite{groot,danielewicz,asakawa} according to classical transport theory. Therefore, qualitatively, the increase of the shear viscosity will reduce the cross section and finally suppress the collisional energy loss. It also should be noted that an explicit expression connecting  the transport parameter $\hat{q}$ governing the radiative energy loss of a propagating parton in QCD plasma with shear viscosity has been derived as
$\frac{T^3}{\hat{q}}\sim \frac{\eta}{s}$ in some recent study\cite{muller99}.

\section{Summary}
\label{summary}
Shear viscosity will modify the distribution functions of the constituents of the QGP. Therefore, it will affect the gluon self-energy, through which the dielectric functions will be influenced. As a result, shear viscosity will have an impact on the polarization energy loss suffered by the fast quark traveling through the QGP. Based on the polarization tensor derived from the viscous chromohydrodynamics, we have evaluated the longitudinal and transverse dielectric functions, through which the polarization energy loss has been investigated. The numerical analysis shows that shear viscosity significantly suppresses the polarization energy loss. The algebraic analysis for integrand in polarization energy loss expression Eq.(\ref{vlossm}) may be helpful for understanding the viscous behavior of it.

The chromohydrodynamic equations  express the macroscopic conservation laws. It should be noted that some dynamical information
will be lost during the derivation from the kinetic theory to the
chromohydrodynamics\cite{manuel1,manuel3,manuel2,jiang2,ramos}. However, in many cases the discrepancies between both approaches of the kinetic theory and the
chromohydrodynamics can be alleviated by using effective parameters as
inputs in the hydrodynamic formalisms\cite{manuel1,ramos}. In chromohydrodynamics, the effective speed sound $c_s$ plays the role of that effective parameter\cite{manuel1,jiang2}. But it recovers part of dynamical information which are relevant to the longitudinal dielectric properties of  plasma, for detailed illustration, please refer to the Appendix of the literature \cite{manuel1}.  Nevertheless, such phenomenological model of the chromohydrodynamics could capture  some correct physics of the QGP\cite{manuel1,manuel3,manuel2,jiang2,ramos}. In view of the difficulty in investigating  the viscous effect on the chromoelectromagnetic properties of QGP in microscopic kinetic theory description, we expect that we could obtain some insight on the physics of the  problem by applying the viscous chromohydrodynamics.

To consistently investigate the viscous effect on the collisional energy loss, in addition to the polarization loss, one should take into account the energy loss in the viscous QGP due to large momentum transfer (individual collisions between incident quark and medium partons).
The systematic field theory way for investigating the collisional energy loss is formulated by Braaten and Thoma by  using the interaction rate
which is related to the damping rate\cite{braaten91a,braaten91b,thoma95}.
The damping rate  in hot viscous QGP has been addressed in a recent literature\cite{sarkar13}, but incorporating a distribution function modified by shear viscosity with a boost invariant expansion formulism  without transverse flow.  To investigate collisional energy loss in the viscous QGP with a consistent framework is an interesting work and deserves further comprehensive study.

After we finished this  work , we noticed that a related work was recently done in \cite{Calzetta14}.

\bigskip
{\bf Acknowledgment}
We would like to extend our gratitude to Guang-You Qin for valuable discussions. B.F Jiang is partly supported   by NSFC under Grant Nos.\ 11147012, 11365008, Defu Hou is supported by NSFC under Grant Nos. 11375070, 11135011 and  11221504 and Jia-rong Li is supported by NSFC under Grant No.11275082.

\bigskip
\appendix
\section{An Alternative derivation of the transverse dielectric function Equation (\ref{dft})}
As discussed in Ref.\cite{ichimaru} and annotated in the nineteenth reference in Ref.\cite{koike} that in addition to the combination ($\varepsilon_L,\varepsilon_T$), the following electric permittivity $\varepsilon$ and the magnetic permeability $\mu_M$ are usually used to describe the electromagnetic properties in plasma. The combination ($\varepsilon_L,\varepsilon_T$) can be related to $\varepsilon$ and $\mu_M$ as following\cite{koike}
\begin{equation}
\varepsilon=\varepsilon_L
\end{equation}
\begin{equation}\label{a2}
\frac{1}{\mu_M}=1+(\frac{\omega^2}{k^2})[\varepsilon_L(\omega,k)-\varepsilon_T(\omega,k)].
\end{equation}
According to (\ref{a2}), one can get
\begin{equation}
\frac{1}{\mu_M}(\frac{k^2}{\omega^2})=(\frac{k^2}{\omega^2})+
\varepsilon_L(\omega,k)-\varepsilon_T(\omega,k),
\end{equation}
and following expression after some algebra,
\begin{equation}\label{mudt}
\varepsilon_T(\omega,k)=(\frac{k^2}{\omega^2})+
\varepsilon_L(\omega,k)-\frac{1}{\mu_M}(\frac{k^2}{\omega^2}).
\end{equation}

In Ref.\cite{jiang2}, in order to investigate the refractive index in viscous QGP, the authors have derived the chromoelectric permittivity $\varepsilon=\varepsilon_L$, ie (\ref{dfl}) and chromomagnetic permeability with chromohydrodynamics
\begin{eqnarray}
\mu_M(\omega,k)
=\frac{1}{1+\frac{\omega_p^2}{k^2}\cdot\frac{1}{1-\frac{\eta}{s}\cdot\frac{k^2}{T\omega}}
+\frac{\omega^2}{k^2}\cdot(\varepsilon(\omega,k)-1)}.\label{vmag}
\end{eqnarray}
Substituting (\ref{vmag}) into (\ref{mudt}), we will arrive
\begin{eqnarray}
\varepsilon_T(\omega,k)&=&(\frac{k^2}{\omega^2})+\varepsilon_L(\omega,k)-(\frac{k^2}{\omega^2})
(1+\frac{\omega_p^2}{k^2}\cdot\frac{1}{1-\frac{\eta}{s}\cdot\frac{k^2}{T\omega}}
\nonumber\\&+&\frac{\omega^2}{k^2}\cdot(\varepsilon(\omega,k)-1))
\nonumber\\&=&
1-\frac{\omega_p^2}{\omega^2}\cdot\frac{1}{1-\frac{\eta}{s}\cdot\frac{k^2}{T\omega}},
\end{eqnarray}
which is coincided with (\ref{dft}).

\section{Derivation of equation~(\ref{18})}
After the relation $m^2_D=3\omega^2_P$ has been adopted\cite{mustafa06}, the longitudinal dielectric function Eq.(\ref{dfl}) can be expressed as
\begin{eqnarray}
\varepsilon_L(\omega,k)
 &=&1+\frac{m_D^2}{k^2}[1-\frac{\omega}{2k}
 (\log|\frac{\omega+k}{\omega-k}|)]
-\frac{4m_D^2}
 {k^2}\frac{\eta\omega}{sT}\nonumber\\&\times&
 \{1-\frac{\omega}{k}
 \log|\frac{\omega+k}{\omega-k}|
 +\frac{\omega^2}{4k^2}
 (\log|\frac{\omega+k}{\omega-k}|)^2\nonumber\\&-&
 \frac{\omega^2}{4k^2}\pi^2\}+ i\{\frac{\pi\omega m_D^2}{2k^3}-\frac{4m_D^2}
 {k^2}\frac{\eta\omega}{sT}\nonumber\\&\cdot&(\frac{\omega}{k}\pi-\frac{\omega^2\pi}{2k^2}
 \log|\frac{\omega+k}{\omega-k}|)\}.
\end{eqnarray}
One can derive the numerator $Nu$ and denominator $De$ of $\rm Im \varepsilon_L^{-1}(\omega,k)$ as following
\begin{eqnarray}\label{nu}
Nu=-\frac{\pi\omega m_D^2}{2k}\{\frac{1}{k^2}-\frac{8\omega}
 {k^2}\frac{\eta}{sT}(1-\frac{\omega}{2k}
 \log|\frac{\omega+k}{\omega-k}|)\},
 \end{eqnarray}
\begin{eqnarray}\label{de}
De&=&[1+\frac{m_D^2}{k^2}(1-\frac{\omega}{2k}
 (\log|\frac{\omega+k}{\omega-k}|))-\frac{4m_D^2}
 {k^2}\frac{\eta\omega}{sT}\nonumber\\
&\times&
 \{1-\frac{\omega}{k}
 \log|\frac{\omega+k}{\omega-k}|
+\frac{\omega^2}{4k^2}
 (\log|\frac{\omega+k}{\omega-k}|)^2\nonumber\\&-&
 \frac{\omega^2}{4k^2}\pi^2\}]^2+[\frac{\pi\omega m_D^2}{2k}\{\frac{1}{k^2}-\frac{8\omega}
 {k^2}\frac{\eta}{sT}\nonumber\\&\cdot&(1-\frac{\omega}{2k}
 \log|\frac{\omega+k}{\omega-k}|)\}]^2.
\end{eqnarray}
Multiplying the numerator $Nu$ and denominator $De$ of $\rm Im \varepsilon_L^{-1}(\omega,k)$ by $k^4$ simultaneously and using the relation $\omega=\textbf{k}\cdot\textbf{v}=kv\cos\theta$, one can get
\begin{eqnarray}\label{nuf}
Nu&=&-\frac{m_D^2\pi v \cos\theta}{2}\cdot\{k^2-k^3\frac{8v\cos\theta}
 {T}\frac{\eta}{s}\nonumber\\&\cdot&(1-\frac{v\cos\theta}{2}
 \log|\frac{1+v\cos\theta}{1-v\cos\theta}|)\},
\end{eqnarray}
and
\begin{eqnarray}\label{def}
De&=&\{[k^2+m_D^2(1-\frac{v\cos\theta}{2}
 (\log|\frac{1+v\cos\theta}{1-v\cos\theta}|))
\nonumber\\&-&k\frac{4m_D^2v\cos\theta}
 {T}\frac{\eta}{s}\times
 \{1-v\cos\theta
 \log|\frac{1+v\cos\theta}{1-v\cos\theta}|
\nonumber\\&+&\frac{v^2\cos^2\theta}{4}
 (\log|\frac{1+v\cos\theta}{1-v\cos\theta}|)^2-
 \frac{v^2\cos^2\theta}{4}\pi^2\}]^2\nonumber\\&+&\frac{m_D^4\pi^2 v^2 \cos^2\theta}{4}[1-k\frac{8v\cos\theta}
 {T}\frac{\eta}{s}\nonumber\\&\cdot&(1-\frac{v\cos\theta}{2}
 \log|\frac{1+v\cos\theta}{1-v\cos\theta}|)]^2\}.
\end{eqnarray}
By defining the following polynomials
\begin{eqnarray}\label{parameter}
&w_1&=\frac{8v\cos\theta}
 {T}\frac{\eta}{s}(1-\frac{v\cos\theta}{2}
 \log|\frac{1+v\cos\theta}{1-v\cos\theta}|),\nonumber\\
 &w_2&=m_D^2(1-\frac{v\cos\theta}{2}
 (\log|\frac{1+v\cos\theta}{1-v\cos\theta}|)),\nonumber\\
 &w_3&=-\frac{4m_D^2v\cos\theta}
 {T}\frac{\eta}{s}\times
 \{1-v\cos\theta
 \log|\frac{1+v\cos\theta}{1-v\cos\theta}|
\nonumber\\&+&\frac{v^2\cos^2\theta}{4}
 (\log|\frac{1+v\cos\theta}{1-v\cos\theta}|)^2-
 \frac{v^2\cos^2\theta}{4}\pi^2\},
 \nonumber\\
 &w_4&=\frac{m_D^4\pi^2 v^2 \cos^2\theta}{4},
 \end{eqnarray}
according to (\ref{nuf}) and (\ref{def}), we can arrive the final expression of $\rm Im \varepsilon_L^{-1}(\omega,k)$ as
\begin{eqnarray}
& &\rm Im \varepsilon_L^{-1}(\omega,k)=-\frac{m_D^2\pi v \cos\theta}{2}\nonumber\\ &\cdot&\frac{k^2-w_1\cdot k^3}{[k^2+w_3\cdot k+w_2]^2+w_4\cdot[1-w_1\cdot k]^2}.
\end{eqnarray}


\end{document}